# Chaotic mixing in a planar, curved channel using periodic slip

P. Garg[1], J. R. Picardo[2] and S. Pushpavanam[2,a]

[1]*Department of Chemical Engineering, Indian Institute of Technology Roorkee, Roorkee, 247 667, India*
[2]*Department of Chemical Engineering, Indian Institute of Technology Madras, Chennai, 600 036, India*

We propose a novel strategy for designing chaotic micromixers using curved channels confined between two flat planes. The location of the separatrix between the Dean vortices, induced by centrifugal force, is dependent on the location of the maxima of axial velocity. An asymmetry in the axial velocity profile can change the location of the separatrix. This is achieved physically by introducing slip alternatingly at the top and bottom walls. This leads to streamline crossing and Lagrangian chaos. An approximate analytical solution of the velocity field is obtained using perturbation theory. This is used to find the Lagrangian trajectories of fluid particles. Poincare sections taken at periodic locations in the axial direction are used to study the extent of chaos. The extent of mixing, for low slip and low Reynolds numbers, is shown to be greater when Dean vortices in adjacent half cells are counter-rotating. Wide channels are observed to have much better mixing than tall channels; an important observation not made for separatrix flows till now. Eulerian indicators are used to gauge the extent of mixing with varying slip length and it is shown that an optimum slip length exists which maximizes the mixing in a particular geometry.

______________________________
[a] Author to whom correspondence should be addressed. Electronic mail: spush@iitm.ac.in, Fax: +91-44-22570509, Ph: +91-44-2254161

## I. INTRODUCTION

Increased focus on microfluidic devices has brought analysis of simple flow fields exhibiting scalar 'chaotic mixing' to the forefront in a variety of fields such as applied dynamical systems and microfluidics[1–5]. Consequently, several 'chaotic mixer' designs have been proposed and examined. We propose a novel design strategy relying on a simple geometry modification leading to chaotic mixing in flow in a curved channel confined between two planes. Dean vortices in curved channels are a classical example of secondary flows transverse to the axial flow, induced by centrifugal force[6,7]. The proposed mixer is composed of a repeating pattern of two alternating half cells; each half cell composed of a curved channel in which the fluid slips at the top or bottom wall, alternatingly. These half cells make up a unit cell, which is repeated periodically. The slip leads to an asymmetry in the geometry which causes the Dean vortices to alternatively grow and shrink. Streamlines in alternate cells, hence exhibit transverse intersections, providing an inherent mechanism for chaotic mixing. The proposed modification to the best of our knowledge is the first which leads to a variation in the location of the separatrix between the Dean vortices.

Ottino and Wiggins[3] give an excellent overview of the different approaches used to design chaotic micromixers. This study is concerned with the design of a passive micromixer i.e. a mixer which uses the geometry of the channel to induce chaotic mixing without any moving parts. The fundamental idea behind passive micromixers is to induce a periodically altered secondary flow transverse to the axial flow, such that streamlines in different sections appear to cross. The symmetry of most passive micromixers implies that the secondary flow is composed of two vortices with a separatrix between them. Such flows are referred to as 'separatrix flows'[8]. Ottino[9] first observed that apparent "streamline crossing", in periodically repeated cells, can be considered as an indicator of Lagrangian chaos. Briefly, the approaches for "streamline crossing" in passive micromixers may be classified as either 'rotation of the vortices'[10,11] or 'movement of the separatrix'[12,13] in sections, which alternate in the axial flow direction.

Dean vortices represent a class of secondary vortices induced by a simple geometry modification i.e. curvature of the channel. Consequently, a lot of work has been devoted to designing micromixers which induce chaotic advection in curved channels[11,14–19]. Aref et al[11] proposed the first design for the same; by taking the channel out of the plane and consequently rotating the Dean vortices[14,20]. Some of the other geometries proposed based on this idea have included: 3 dimensional L-shaped bends[15], F-shaped units[16], square waves[14], spirals[18] and so on. These designs are based on the same principle: rotation of Dean vortices. The complex geometry modifications required render easy integration into lab on chip devices unfeasible. An alternative approach has been to exploit the asymmetry caused at high Reynolds numbers due to the bifurcation of the two vortices to the four vortex pattern[17]. This approach requires high Reynolds numbers and curvature ratio and hence, is unsuitable for micromixers.

However, all these approaches preserve the symmetry of the flow about the centerline (in the direction of the centrifugal force) of the cross-section. Consequently, the separatrix between the Dean vortices does not move and is located at the centerline. The alternative approach i.e. 'movement of the separatrix', thus, has not been explored for curved micromixers, despite it having been used in the design of micromixers based on other geometry modifications[4,13,21]. The general strategy here is to vary the location of separatrix between the secondary vortices from one section to the next by modifying the geometry, which leads to the requisite transverse "streamline crossing". The flow field can then be expected to lead to chaotic mixing. The justification for this hypothesis is based on Linked Twist Maps (LTMs), which have been shown to be ergodic[5,8]. The inherent symmetry of the Dean vortices at first sight might lead one to conclude that varying the location of the separatrix in a curved channel is not possible. This is probably why this mechanism has never been explored for curved channels. However, from recent work[22], we have discovered that the location of the separatrix depends (amongst other things) on the maxima of the axial velocity. This suggests an easy way to introduce asymmetry in the Dean vortices: by making the axial velocity profile asymmetric. In the design discussed in this paper, this is achieved by introducing slip alternatingly at the top and bottom wall of the channel in a periodic manner. The maxima of the axial velocity consequently moves near the top (or bottom) wall. Thus, the location of the separatrix varies between the two sections, leading to streamline crossing and Lagrangian chaos. Although, secondary vortices rely on inertia, chaos is observed for finite but low Reynolds numbers as the asymmetry in the proposed design is independent of inertial effects.

Our design is made feasible by recent advances in the study of slip in microchannels[23,24]. Since, drag reduction is a major challenge in microchannels, the ability to induce slip has become increasingly important. Recent work on super-hydrophobic surfaces has shown that slip lengths of 25 μm and drag reduction of 40% can be achieved[25]. Rothstein[26] has a recent detailed review of this aspect. Despite the fact that enhanced control over "slip" is possible in microchannels, not many devices exist which exploit it for purposes other than simple drag reduction[27]. Our design marks a first step in this direction. Additionally, the slip is expected to facilitate drag reduction, both for the axial flow as well as the secondary Dean flow[28,29]. Thus, our design combines two optimum features of microchannels to obtain chaotic mixing: asymmetric secondary Dean vortices at low Reynolds numbers and drag reduction due to slip. However, in the present study we do not assume a particular cause of slip, so the consequent analysis is independent of the mechanism of slip.

Struman et al[8] proposed that counter-rotating Linked Twist Maps (LTMs) on the plane have better mixing properties than co-rotating LTMs for a wider range of parameters. A similar observation was made by Kang et al[15], who found out that, for their mixer design, mixing is better when the Dean vortices are counter-rotating in successive half cells. Thus, motivated we study mixing in two different unit cell geometries which cover both the preceding scenarios: circlet and serpentine. We demonstrate that the response to introduction of asymmetry is qualitatively different in the two cases. This has practical implications for the optimization of the two designs and implies that the underlying symplectic maps are qualitatively different. Both the geometries are confined between two parallel planes. This is an immense simplification over existing designs of chaotic mixers in curved channels.

Techniques used for quantifying chaos and optimizing micromixers, range from calculating Lyapunov exponents to variance of an initial distribution of tracers. These methods require repeated computations of the Lagrangian paths of fluid particles, making it computationally intensive to efficiently and thoroughly span the parameter space[30]. To overcome this difficulty, Struman and Wiggins[31] have introduced the concept of Eulerian indicators, which may be used to predict the quality of mixing based on the velocity field alone. Consequently, for optimization of the mixer in the 2x2 {(serpentine, circlet) x (slip length, aspect ratio)} parameter space, we use Eulerian indicators in conjunction with Poincare sections. To the best of our knowledge, the variation with aspect ratio has not been studied in the literature before for separatrix flows.

The paper is organized in the following manner. In section II, the design geometry and the velocity field in the mixer is elucidated. The infinite time transport properties of the system are studied in Sec. III using Poincare sections. In Sec. IV, we study the change in the extent and nature of chaotic mixing across the parameter space with the aim of narrowing down the parameter space for optimization. In Sec. V, we briefly discuss finite time transport, in the parameter space of interest identified in the previous section. We end by summarizing the key results and discussing their significance in section VI.

## II. MIXER DESIGN AND VELOCITY FIELD

The basic geometry of the mixer is a curved channel, with a rectangular cross section, composed of sections, half cells, with slip at the top and bottom wall, alternatingly. The section with slip at the top is referred to as half cell 1, while that with slip at the bottom wall as half cell 2. Each half cell spans an azimuthal extent of $180^0$. The two half cells together form a unit cell, which is repeated periodically to form the mixer. Two different arrangements of these alternative sections are considered throughout the paper. If the direction of centrifugal force is same in the two sections, it is referred to as 'Circlet' (Fig **1** (a)); if the direction is opposite, it is referred to as 'Serpentine' (Fig **1** (b)). The top view of the geometry of each unit cell for both configurations is depicted in Fig **2**. The curvilinear coordinate system *(x, y, θ)* used in the study is depicted in Fig 3. The origin of the *x-y* plane (*O'*) is located at the center of a cross section of the channel. The azimuthal *θ* coordinate determines the axial position of the cross section.

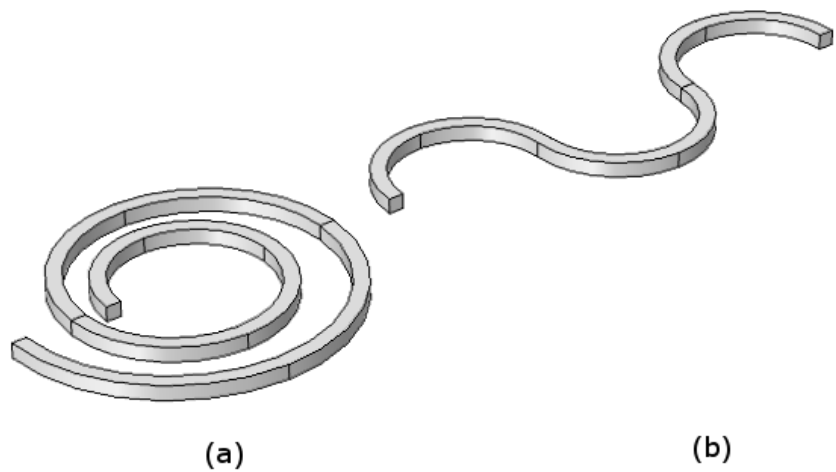

Fig 1 The two different mixer designs (a) Circlet (b) Serpentine

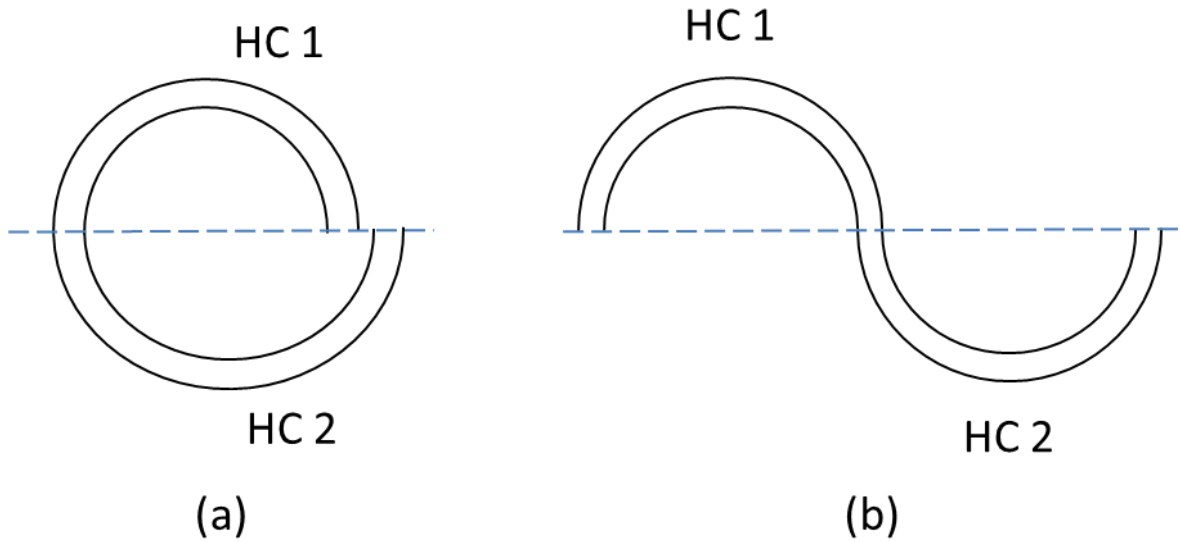

Fig 2 Top view schematic of the (a) circlet (b) serpentine unit cells. HC1 and HC2 refer to half cell 1 and half cell 2, respectively. Half cell 1 has slip at the top wall, while half cell 2 at the bottom wall.

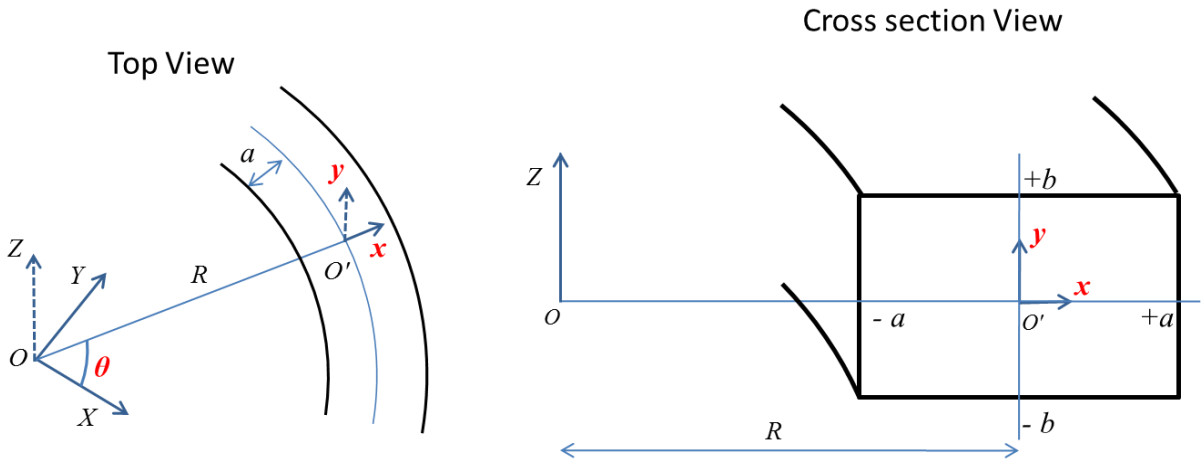

Fig 3 The curvilinear co-ordinate system used in this work.

The asymmetry of the Dean vortices in the x-direction in channels of finite curvature or at high Reynolds numbers can also lead to chaotic mixing[17]. To ensure that chaotic mixing is not caused by these effects, we only study the small curvature ratio and Reynolds number regime. This ensures that any asymmetry is due to a change in the location of the separatrix.

The flow is assumed to be steady and fully developed along the axial direction within each half cell. Hence, the velocity field in each half cell is assumed to be independent of the preceding cell and switches discontinuously at the boundary between cells. This assumption enables us to obtain an analytical solution for the velocity field. This simplified velocity field captures the essential features of the geometry, while simplifying subsequent computations and enabling a thorough parametric study of the system. The quasi, fully-developed assumption has been used several times in the literature to model chaotic mixers [9,10,11]. Further, the development length for curved channels has been shown to be of the order of the width of the channel at low Reynolds numbers[32,33]. Hence, the quasi, fully-developed assumption is justified for the small curvature regime investigated in this work.

The non-dimensional parameters used in this work are defined as:

$$\text{Re} = \left(\frac{1}{4}\left(-\frac{1}{R}\frac{\partial P}{\partial \theta}\right)\right)\frac{\rho a^3}{\mu^2} \qquad \lambda = \frac{a}{b} \qquad \varepsilon = \frac{a}{R} \qquad k = \frac{\beta}{h} \qquad L = \frac{\Theta}{\varepsilon}$$

where Re is the Reynolds number (defined on the average velocity, $v_{avg} = \left(\frac{a^2}{4\mu}\left(-\frac{1}{R}\frac{\partial P}{\partial \theta}\right)\right)$, in a straight channel without slip), $\lambda$ is the aspect ratio (i.e. ratio of the width $(a)$ to the height $(b)$ of the cross section), $\varepsilon$ is the curvature ratio of the channel (i.e. ratio of the width $(a)$ to the radius of curvature $(R)$) and $k$ is the non-dimensionalised Navier slip length which determines the extent of slip, $L$ is the ratio of the length of one half cell $(\Theta R)$ to the width of the channel $(a)$. Unless stated otherwise $\varepsilon = 0.1$ and $\Theta = 180^o$ for each half cell. We note that for the circlet geometry $L$ must vary with each half cell i.e. it must be a spiral. However, this change will not affect the qualitative results discussed as $L$ only determines the non-dimensionalised length of a half cell. Therefore, we neglect the variation of $L$ between different half cells for the circlet geometry, implying we ignore the effects due to the spiral nature of the geometry. Further, in terms of non-dimensional parameters, the quasi-fully developed assumption requires $L \gg 1 \Rightarrow (\varepsilon/\Theta) \ll 1$. We see that this holds in the regime studied.

The domain perturbation method is used to find an asymptotic solution of the governing Navier-Stokes equations for small curvature ratio and Reynolds numbers. Details of the solution procedure and expressions for the velocity field are given in the Appendix.

We first examine the velocity in the unit cell of the circlet geometry. Fluid flowing through a curved channel experiences a centrifugal force towards the outer part of the channel, which sets up a helical flow field. The flow field may be analyzed by decomposing it into an axial flow and a secondary transverse flow in the cross section, composed of two vortices, known as Dean vortices [6,7]. As a first approximation (at $O(\varepsilon^0)$), the velocity in a curved channel is the same as that in a corresponding straight channel. Fig 4 (a) shows the axial velocity for no slip at either wall. The velocity profile is symmetric and the maxima is located at the center. When we introduce slip at the top (bottom) wall i.e. half cell 1 (2), the maxima moves up (down), as seen in Fig 4 (b) (Fig 4 (c)), due to decreased tangential stress at the top (bottom) wall. These contours are obtained by represent the axial velocity, given by $w_0$ of Eqn. A7, Appendix.

The secondary vortices represent the correction due to centrifugal force at $O(\varepsilon^1)$. This yields $u_1$ and $v_1$ in Eqn. A5, Appendix and Eqn. A6, Appendix. Fig 5 (a) shows the Dean vortices, when there is no slip at either wall. Fig 5 (b) shows the Dean vortices for the half cell 1 i.e. slip at the top wall, where the separatrix has been shifted upwards as compared to the case when there is no slip at either wall (Fig 5 (a)). The separatrix between the Dean vortices lies at the location of the maxima of centrifugal force, where the fluid tends to move towards the outer part of the channel. The centrifugal force is in turn dependent on the axial velocity (Eqn. A8, Appendix). Clearly, it follows that an asymmetric axial velocity profile leads to a change in the location of the separatrix. The location of the separatrix is also affected by the lower drag on the vortex slipping at the wall. However, this is an $O(\varepsilon^1)$ effect, the implications of which are discussed in detail later.

The Dean vortices, for the half cell 2 of the circlet geometry are shown in Fig 5 (c). As noted above, the maxima of the axial velocity moves down in this half cell (Fig 4 (c)). Hence, as expected, the separatrix has moved downwards in Fig 5 (c). We note that the velocity profile in the second half cell is the reflection in the x-axis of the velocity profile in the first half cell, for the circlet geometry.

The secondary, circulatory flow in the two half cells of the serpentine geometry is shown in Fig 6. Half cell 1 of both the geometries are identical, hence the velocity fields in them are the same. (See Fig 6(a) and Fig 5 (b)) The velocity field in half cell 2 of the serpentine geometry may be inferred from its relation to half cell 1. For the serpentine geometry, the direction of rotation of the secondary vortices reverses from half cell 1 to half cell 2 as the direction of centrifugal force reverses. This implies the velocity in half cell 2 is a reflection in the y-axis of half cell 1. This is in addition to the change due to slip at different walls, which implies a reflection in the x-axis. Therefore, the velocity in the second half cell is an 180° rotation about the origin of the velocity in the first half cell. Finally, we note that the axial velocity in the unit cell of the serpentine geometry is the same as that in the circlet geometry (Fig 4 (b)-(c)), while the difference in the Dean vortices can be noted by comparing Fig 5 (b)-(c) with Fig 6 (a)-(b).

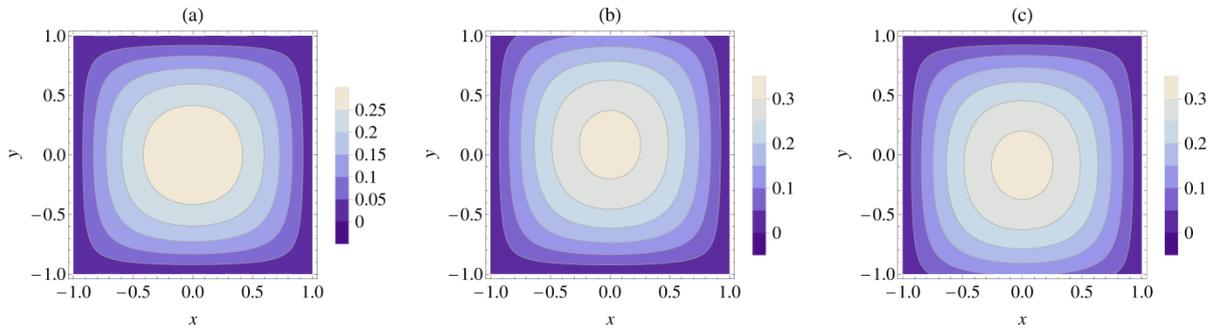

Fig 4 Contour plots of the axial velocity for (a) no slip (k=0) (b) slip at the top wall (k=0.2) (c) slip at the bottom wall (k=0.2).

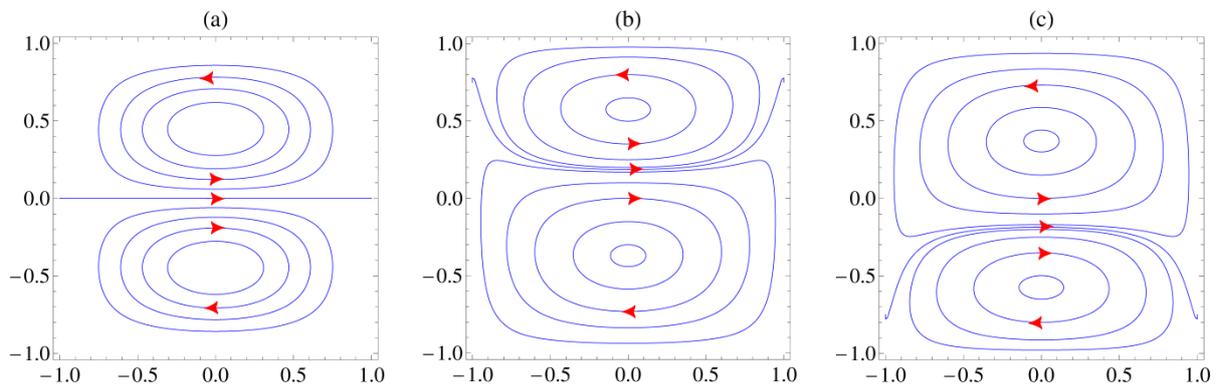

Fig 5 Streamlines projected onto the *x-y* plane for: (a) no slip (k=0) (b) Circlet half cell 1: slip at the top wall (k=0.2) (c) Circlet half cell 2: slip at the bottom wall (k=0.2). The other parameters are: $Re = 30, \lambda = 1$

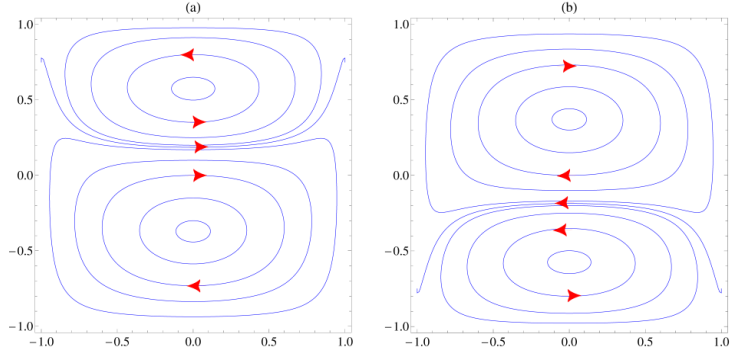

Fig 6 Streamlines projected onto the *x-y* plane for (a) serpentine half cell 1: slip at the top wall (k=0.2) (b) serpentine half cell 2: slip at the bottom wall (k=0.2). The other parameters are: $Re = 30, \lambda = 1$

## III. INFINITE TIME DYNAMICS

The flow field described above is a 2+1 dimensional, quasi fully-developed, flow field, which changes discontinuously from one half cell to the next. The equations of motion of the passive tracer particle are given by:

$$\frac{dx}{dt} = L\varepsilon u_1 = \Theta \frac{\partial \psi(x,y)}{\partial y} \tag{1}$$

$$\frac{dy}{dt} = L\lambda\varepsilon v_1 = -\Theta\lambda \frac{\partial \psi(x,y)}{\partial x} \tag{2}$$

$$\frac{d\theta}{dt} = w_0 = f(x,y) \tag{3}$$

(The expressions for $w_0$ (axial velocity) and $\psi$ (stream function) are given in the Appendix, A9 and A10 respectively.)

Eqn. (1) and Eqn. (2) form a 1 degree of freedom periodic Hamiltonian system. Following the standard practice in the literature on chaotic mixers, we study the infinite-time dynamics of the system by examining the Poincare sections at the end of every unit cell. The system is, thus, reduced to a two-dimensional symplectic map with one degree of freedom. The axial velocity is dependent on the location in the x-y plane; hence different particles take different times to reach the same cross section. However, while constructing the Poincare map, it is assumed that each particle has sufficient time to reach the cross section.

Let $M_1^c$ and $M_2^c$ denote the map from the entry to the exit of the first half cell and second half cell of the circlet unit cell, and $M_1^s$ and $M_2^s$ for the serpentine unit cell, respectively. The four mappings are found by numerically integrating equations (1)-(3) using the inbuilt NIntegrate (numerical integration) function of Mathematica 9.0, which uses a fourth order Runge-Kutta method. The initial points are chosen such that the full structure of the Poincare sections can be found, and hence, the initial distribution varies for each case. The points are tracked for 1000 sections, to ensure that we are studying infinite-time dynamics of the system.

Let, $H^c$ and $H^s$ denote the maps from the entry to the exit of the circlet and serpentine unit cell, respectively.

$$H^c = M_2^c \circ M_1^c \qquad\qquad H^s = M_2^s \circ M_1^s$$

We consider the case with no slip $(k=0)$ as the unperturbed map, while the introduction of slip $(k \neq 0)$ is treated as a perturbation. However, we note that this perturbation need not be small.

We first discuss the circlet geometry. For $k=0$, $H^c$ reduces to flow in a curved channel with no-slip at all the walls as shown by:

$$H^c = M_2^c \circ M_1^c = M_1^c \circ M_1^c = (M_1^c)^2$$

It is instructive to look at the structure of the unperturbed unit cell maps $(k=0)$: $H^c$ has two saddle points at $x = \pm 1, y = 0$ with one heteroclinic orbit joining them: the separatrix. All other points on the wall are fixed points, while all the points inside the domain lie on periodic orbits around two elliptic fixed points, one each in the top and bottom half. The above can be inferred from Fig 7 (a). With the introduction of slip, we see that both chaotic and ordered regions exist in the Poincare sections (Fig 7 (b)-(d)). The size of these regions changes with the system parameters and is studied systematically in the next section.

For the serpentine geometry, $H^s$ is the identity map, when $k=0$, since:

$$H^s = M_2^s \circ M_1^s = (M_1^s)^{-1} \circ M_1^s = I$$

where $I$ is the identity map, implying that all the points are exactly mapped back to their original positions. From Fig 8 (a), we may infer that all the points inside the domain are fixed points for unperturbed $H^s$. Mixing in the serpentine unit cell without slip (unperturbed $H^s$) is much worse than that of circlet unit cell (unperturbed $H^c$), since all points are mapped back to their original positions after one unit cell. For the perturbed case $(k \neq 0)$, we find that the serpentine geometry also exhibits chaotic regions. This can be seen from the Poincare sections in Fig 8 (b)-(d).

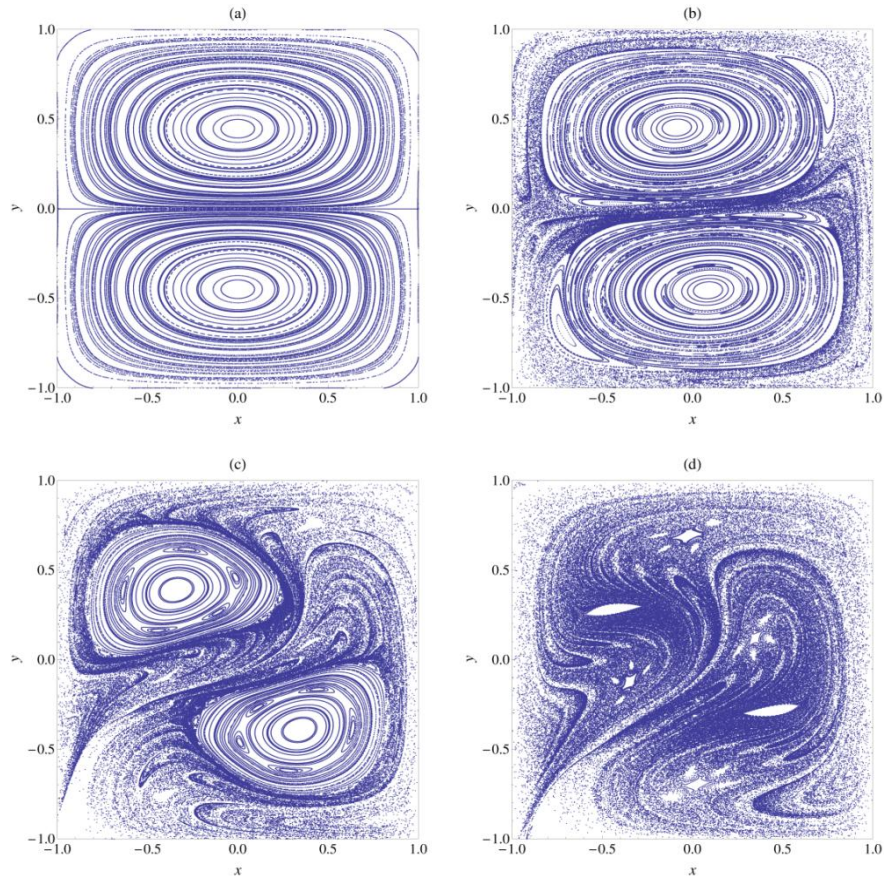

Fig 7 Poincare Sections for the circlet geometry for varying slip length: (a) $k=0$ (b) $k=0.05$ (c) $k=0.25$ (d) $k=0.45$. $Re=15, \lambda=1$ are constant for all the cases. The initial points are chosen such that the full structure of the Poincare sections is shown and the points are tracked for 1000 sections.

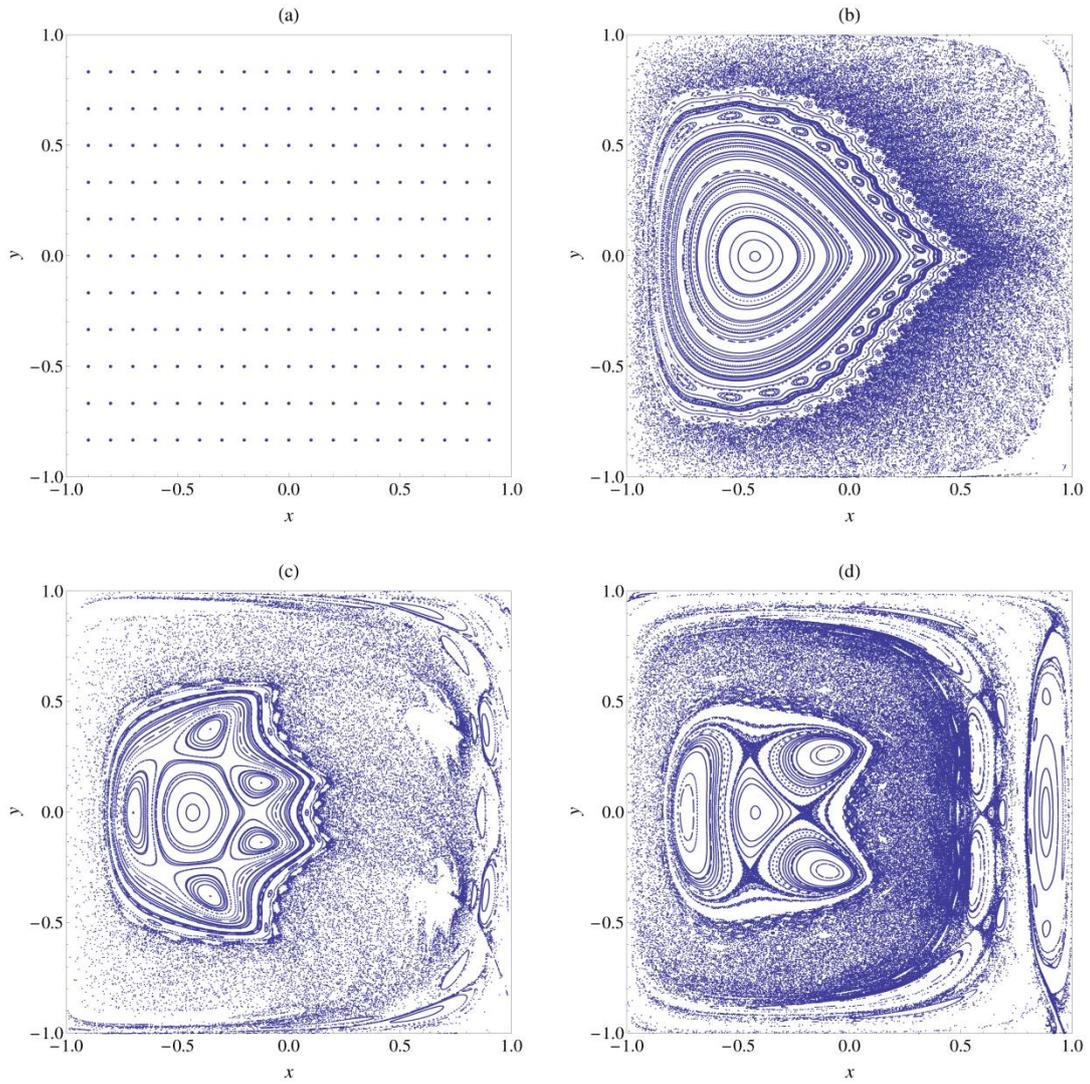

Fig 8 Poincare Sections for the serpentine geometry for varying slip length: (a) $k=0$ (b) $k=0.05$ (c) $k=0.25$ (d) $k=0.45$. $Re=15, \lambda=1$ are constant for all the cases. The initial points are chosen such that the full structure of the Poincare sections is shown and the points are tracked for 1000 sections.

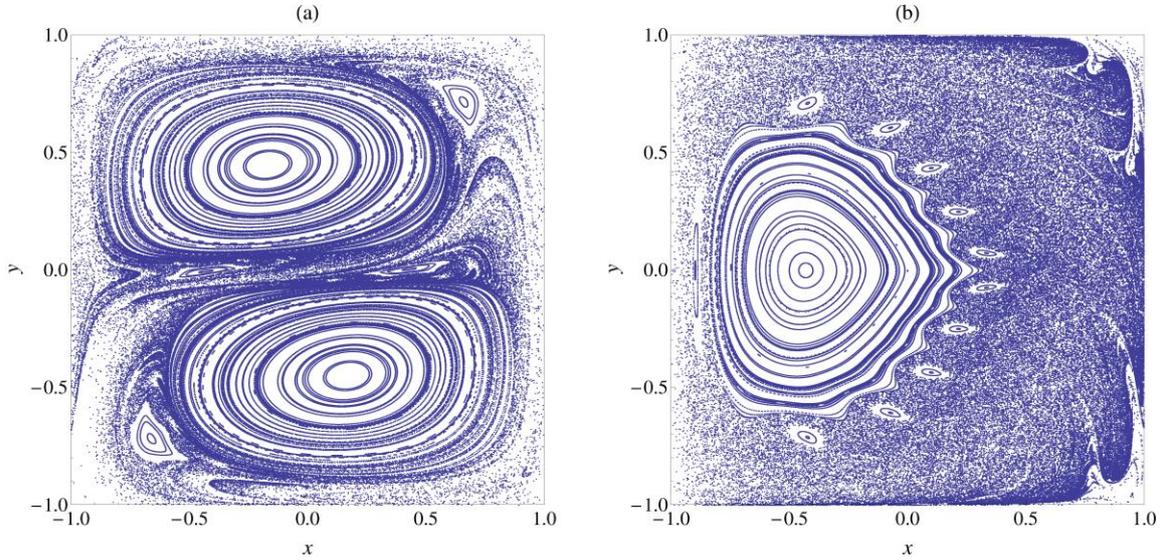

Fig 9 Comparison of (a) circlet and (b) serpentine geometries for parameter values $Re = 15, \lambda = 1, k = 0.1$.

The introduction of periodic slip causes the separatrix to move and the streamlines in the different unit cells cross. McIlhany and Wiggins[34] and Struman and Wiggins[31] study a simplified kinematic velocity field, where the position of the separatrix varies between two half cells, and show that the system is chaotic. Further, it is conjectured in Wiggins and Ottino[5] that the mathematical results demonstrating the ergodic property of Linked Twist Maps (LTMs) strongly suggest that separatrix flows possess ergodic mixing properties. These observations combined with the existence of several chaotic mixers in the literature based on separatrix flows, suggest that chaotic mixing in the proposed geometry is highly likely. This is borne out by the results shown in Fig 7 and Fig 8.

In both Fig 7 and Fig 8, we can see the existence of both regular as well as chaotic regions, which is a characteristic of Hamiltonian systems. The response of the two maps ($H^c$ and $H^s$) is qualitatively different to perturbations. This can be inferred by comparing Fig 7 and Fig 8. For the circlet geometry, most of the periodic orbits from the unperturbed map survive and form the boundary of the ordered region (Fig 7). However, for the serpentine geometry, we see that the fixed points form a number of unstable periodic orbits, which disintegrate with increasing perturbation leading to an increase in the size of the chaotic region (Fig 8). The periodic orbits are usually referred to as KAM tori. The largest of such KAM tori forms a boundary between regions with regular and chaotic motion. Further, KAM tori act as barrier across which fluid particles cannot be transported. Thus, their presence implies that mixing does not occur over the entire domain. The transport in Hamiltonian maps (even in regions which are chaotic) has many subtle features. We refer the reader to the vast literature for further details[35–37]. The Poincare sections in Fig 7 and Fig 8 also show self-similarity and symmetry, a characteristic feature of Hamiltonian maps[38,39]. All these properties have been extensively studied, and their implications for mixing, although an interesting problem is beyond the scope of present work.

From Fig 9, we observe that the serpentine mixer exhibits a larger chaotic region and hence has better mixing properties at low Reynolds numbers and slip length. We can also observe that only one KAM torus is present for the serpentine geometry (Fig 9 (b)), as opposed to the two for the circlet geometry (Fig 9 (a)). It has been shown rigorously that counter-rotating linked twist maps (LTMs) on the plane have better mixing properties than co-rotating LTMs[5,8]. This result implies that the serpentine geometry should be more chaotic. However, as noted by Sturman et al.[8] several assumptions made for LTMs are not valid for separatrix flows. Thus, a rigorous investigation of mixing in the serpentine versus the non-serpentine geometry requires an extension of the LTM theorems to separatrix flows.

**IV. OPTIMIZATION**

The parameters which characterize the flow field are: Reynolds number $(\text{Re})$, aspect ratio $(\lambda)$ and slip length $(k)$, for both the mixer geometries. Reynolds number determines the strength of the circulatory flow field. We restrict ourselves to low Reynolds numbers $(\text{Re} < 30)$ so that no asymmetry is induced due to inertia. In this regime, increasing Reynolds number increases the area of the chaotic region. The slip length $(k)$ determines the location of the separatrix and hence the transversality of the intersection between the streamlines. Aspect ratio $(\lambda)$ affects both the strength of the circulatory flow as well as the location of the separatrix.

Optimization of micromixers has traditionally been done by calculating either Lyapunov exponents or variance of an initial distribution of tracer particles. However, both these approaches require the knowledge of the particle paths, and hence are computationally intensive. Recently, a new approach to optimize micromixers based on the Eulerian characteristics of the velocity field has been put forward by Wiggins and co-workers[31,34,40,41].

They have demonstrated that the approach works well for separatrix flows, and can help one narrow down the parameter space for optimum designs for which further numerical studies can be performed. Here, "streamline crossing" is considered essential for chaotic mixing, and an Eulerian indicator, $\alpha$ is defined to quantify the mixing. Following Struman and Wiggins [31], we define $\alpha(x,y)$ as the angle between the streamlines at $(x,y)$.

$$\alpha(x,y) = cos^{-1} \frac{V_1.V_2}{|V_1||V_2|} \qquad \text{where } V_i = (u_i, v_i)$$

$\alpha(x,y)$ is further restricted to lie between $0^o$ to $90^o$, so the directions of the velocity vectors do not matter. This ensures that if the streamlines cross twice, but in opposite directions, their contribution to $\alpha(x,y)$ is not nullified. The average of $\alpha(x,y)$ over $-1 \leq x \leq 1, -1 \leq y \leq 1$ is denoted by $\bar{\alpha}$ and is the desired Eulerian indicator[31]. Since, the slip length $(k)$ determines the location of the separatrix, $\bar{\alpha}$ is particularly suited to study the changes in the system with varying slip length. (Fig 10).

*Variation of Slip Length*

Let us first consider the effects of varying slip length for a fixed aspect ratio. Increasing slip length, leads to a greater shift of the maxima of the axial velocity, hence one vortex grows, while the other shrinks. At first, this leads to an increase in "streamline crossing" and hence $\bar{\alpha}$ and thus the system can be expected to become more chaotic. However, as the vortex size keeps increasing, the other almost completely shrinks, leaving only one vortex to occupy the cross-section. Hence, overlap between streamlines reduces at large slip, thus implying an optimum slip length that maximizes chaotic mixing. This can be seen from Fig 10 (a)-(f), which shows the value of the parameter $\bar{\alpha}$ as the slip length is varied for different aspect ratios.

We can see from Fig 10 (e) and (f) that for $\lambda = 3, 4$, the curve exhibits two maximas. It can be seen that $k = 0.05$ has a greater value of $\bar{\alpha}$ than $k = 0.1$, for $\lambda = 3$, implying that lower slip leads to more efficient mixing (Fig 10 (e)). This counter-intuitive result may be explained by noting that the location of the separatrix is also governed by the relative strength of the vortices. Consider the case of slip at the top wall. Since the relative area of slip increases for wider channels, hence the strength of the top vortex also increases. Consequently, the separatrix is pushed downwards. However, the asymmetry of the axial velocity opposes this effect, causing the separatrix to move upwards. At large aspect ratios $(\lambda \geq 3)$ and low slip, the asymmetry of the axial velocity is reduced and the effect of the vortex strength dominates. Hence, the top vortex is able to push the separatrix downwards, below the centerline, at low slip lengths (Fig 11 (a)). However, since this is an $O(\varepsilon^1)$ effect, with increasing slip the effect due to the asymmetry of the axial velocity ($O(1)$) dominates. Therefore, as the slip length is increased the separatrix moves upwards (Fig 11 (b)-(c)). Hence, the separatrix moves towards the centerline with increasing slip. The relative overlap between the streamlines thus reduces initially. This leads to the first hump in $\bar{\alpha}$ observed in Fig 10 (e). Fig 12 shows that the Poincare section for $k=0.05$ is indeed more chaotic than $k=0.1$ for $\lambda=3$. Fig 10 (f) shows that this effect is more pronounced for $\lambda=4$, thus validating the preceding argument. We remark that this non-intuitive result would have been especially hard to locate with the traditional Lagrangian methods used to characterize mixing in chaotic flows. This example, hence, clearly shows the important role that the Eulerian indicators, pioneered by Wiggins and co-workers, can play in the optimization of chaotic mixers.

$\bar{\alpha}$ is independent of the Reynolds number and the type of unit-cell (serpentine or circlet). Trends predicted by the Eulerian indicator, $\bar{\alpha}$, have been seen to be true for both serpentine and circlet mixer geometries, by observing the Poincare sections for a selected number of cases at low values of Reynolds numbers $(Re \ll 20)$. For relatively high Reynolds number, $(Re \sim 20, \lambda=1)$ for the serpentine mixer, we found that $\bar{\alpha}$ does not predict the trend accurately. Increasing values of the slip length exhibits lesser chaos, contrary to the trend predicted by $\bar{\alpha}$.

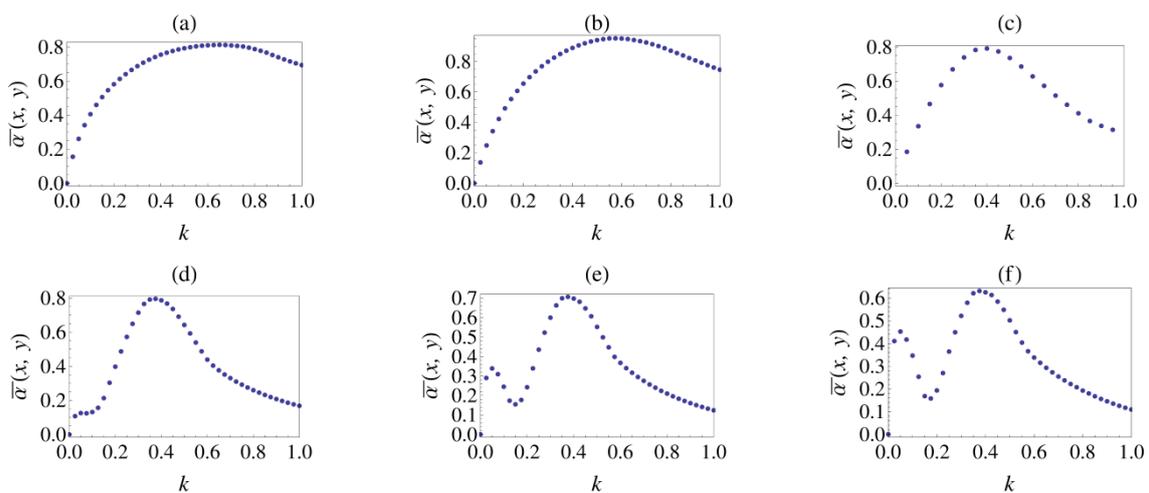

Fig 10 Variation of the Eulerian indicator versus the slip length for different aspect ratios: (a) $\lambda=0.33$ (b) $\lambda=0.5$ (c) $\lambda=1$ (d) $\lambda=2$ (e) $\lambda=3$ (f) $\lambda=4$

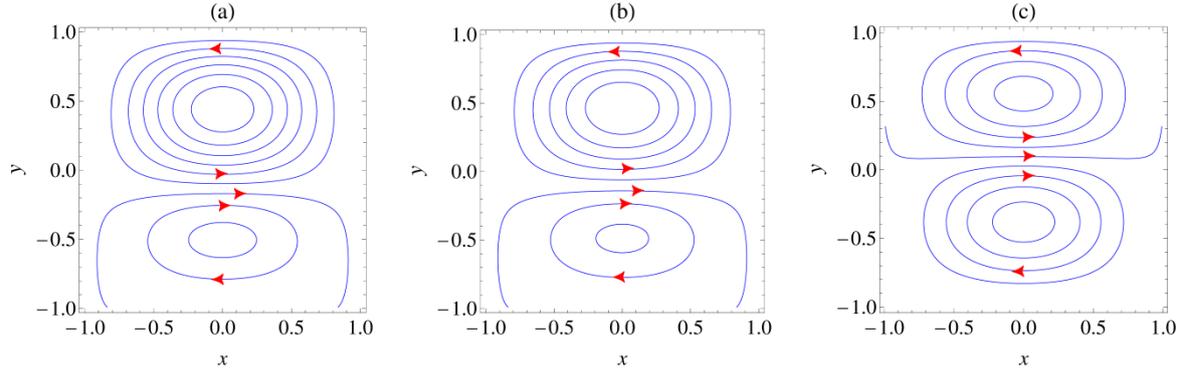

Fig 11 Location of the separatrix with varying slip in wider channels has a non-monotonic dependence on the slip length. (a) $k=0.05$ (b) $k=0.1$ (c) $k=0.25$

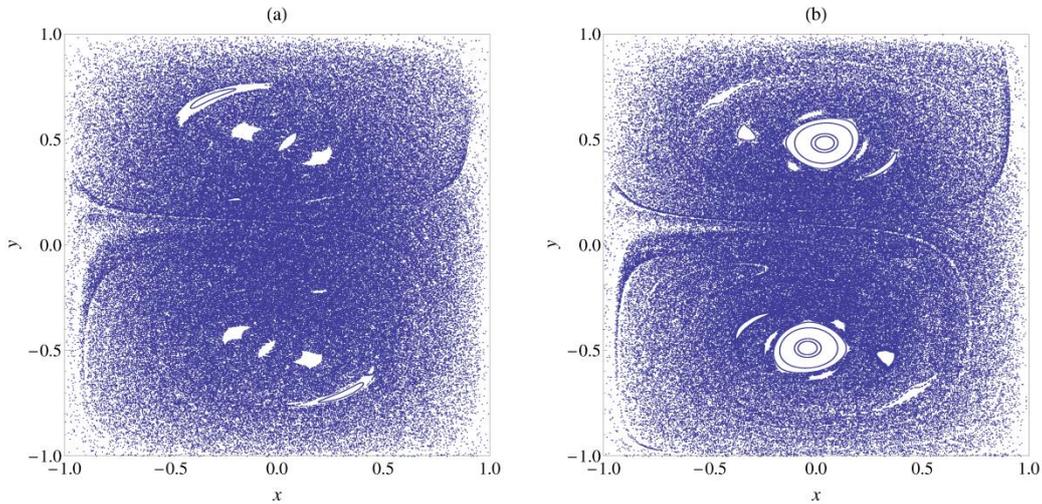

Fig 12 An illustration of the non-intuitive phenomena that less slip can lead to greater chaos in wider channels. Poincare sections for: (a) $\text{Re}=15.4, \lambda=3, k=0.05$ (b) $\text{Re}=15.4, \lambda=3, k=0.1$

*Variation of Aspect Ratio*

The dependence of the mixing efficiency of chaotic mixers based on separatrix flows on the aspect ratio of the channel has not been investigated in the literature before. Considering variations with the aspect ratio provides an additional parameter in the problem, which as we show below, is crucial for optimization.

From Fig 10 (c) and (d), we note that the value of $\bar{\alpha}$ is less for $\lambda=2$ as compared to $\lambda=1$. However, the extent of chaos increases clearly for the wider channel as can be seen from the Poincare sections in Fig 13 and Fig 14. This anomaly can be explained by noting that changes in aspect ratio while affecting the location of the streamlines, however also affect the strength of the secondary circulations. Since, the second affect is not captured by $\bar{\alpha}$; it cannot be used for a comparison between channels of different aspect ratios. (In comparisons between different aspect ratios, the cross-sectional area and the pressure drop are kept constant.)

From Fig 13 and Fig 14, we see that wider channels have much better mixing for both the geometries. For wider channels the relative area of slip is more which implies that the relative area of "streamline crossing" increases, thus, the extent of chaos is expected to increase. We note that for the serpentine geometry, on increasing the aspect ratio, two KAM tori are present instead of one (Fig 14).

Fig 15 shows that for taller channels no chaotic regions are present for $k = 0.1$. Taller channels imply crossing of the streamlines over a lesser area, hence they have poorer mixing properties for both the serpentine and circlet geometries.

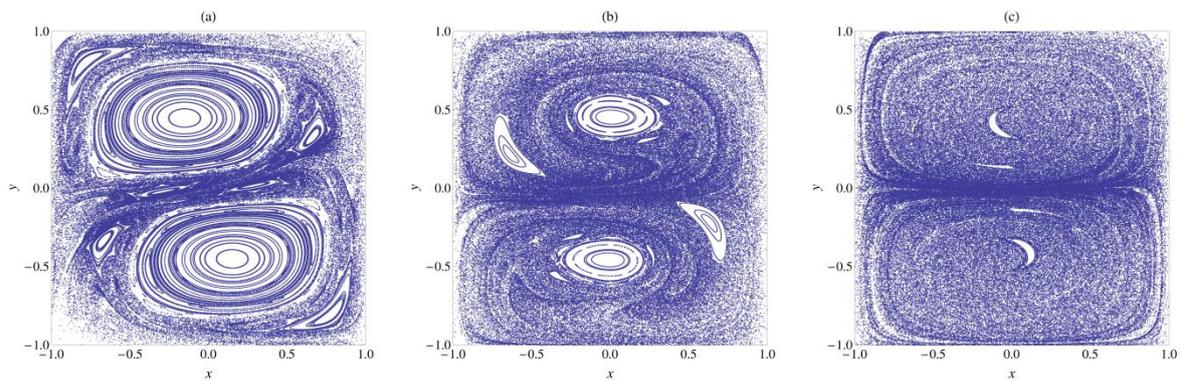

Fig 13 Variation in the area of the chaotic region for the circlet geometry for wider channels (a) $Re = 10, \lambda = 1, k = 0.2$ (b) $Re = 10, \lambda = 2, k = 0.2$ (c) $Re = 10, \lambda = 3, k = 0.2$.

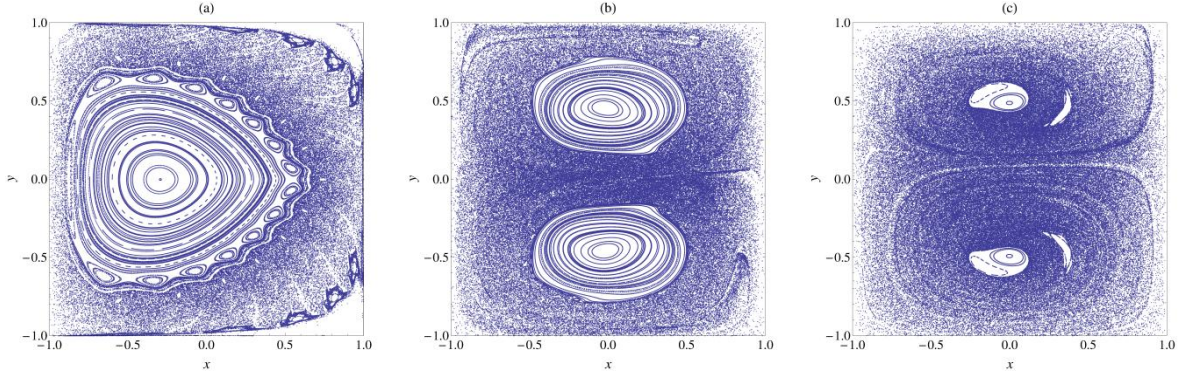

Fig 14 Variation in the area of the chaotic region for the serpentine geometry for wider channels (a) $Re=10, \lambda=1, k=0.1$ (b) $Re=10, \lambda=2, k=0.1$ (c) $Re=10, \lambda=3, k=0.1$.

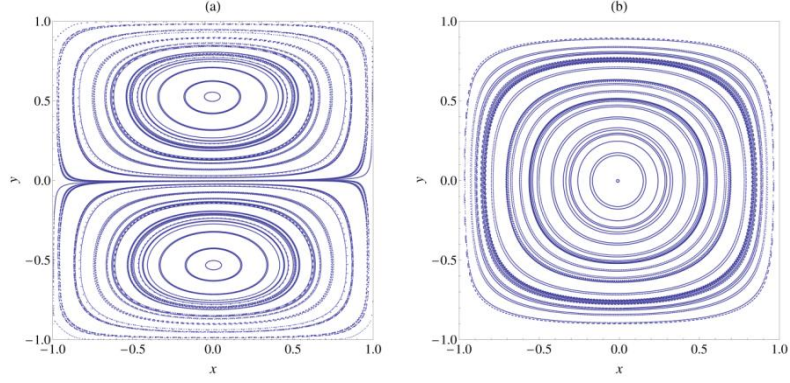

Fig 15 Poincare sections for tall channels: (a) Circlet geometry ($Re=10, \lambda=0.5, k=0.2$) (b) Serpentine geometry ($Re=10, \lambda=0.5, k=0.1$)

## V. FINITE TIME TRANSPORT

So far, we have looked at the infinite time dynamics of the system. To understand the underlying geometrical features that govern mixing, we look at the transport of particles initially in the top half of the channel after a finite number of unit cells. Further, in experiments the transport of tracers from one half of the channel to the other may be observed. Thus, motivated we study the transport of particles initially in the top half of the channel after a finite number of unit cells. For the unperturbed geometries $(k=0)$, the particles in the top half are not transported to the bottom half. For the circlet geometry mixing takes place; however only in the upper half of the channel i.e. no particle crosses the separatrix ((Fig 16(a)). In contrast, for the unperturbed serpentine geometry, all the particles stay at their initial position ((Fig 16(b)).

With the introduction of slip, the underlying geometrical structures governing transport become different. These are shown for the three qualitatively different cases: Circlet (Fig 17), Serpentine (Fig 18) and Serpentine-wide $(\lambda = 3)$ (Fig 19). The initial tracer locations for all cases can be seen in Fig 16(b). Clearly, the transport from the top half of the channel to the bottom half is more efficient for the serpentine geometry (Fig 18 (b)), after 3 unit cells, as compared to the circlet geometry (Fig 17(b)). Hence, the extent of mixing follows the qualitative discussion above. From Fig 18 and Fig 19, we can compare the transport of particles in serpentine geometry of different aspect ratios. The transport of particles from the upper to lower half is seen to be qualitatively different for the wider channel. We note that from these figures only a qualitative comparison can be made between the different cases. A systematic study of transport at finite times can be carried out using lobe dynamics i.e. the study of the geometrical structure formed by the intersection of the stable and unstable manifolds of hyperbolic fixed points[37]. However, it lies beyond the scope of present work.

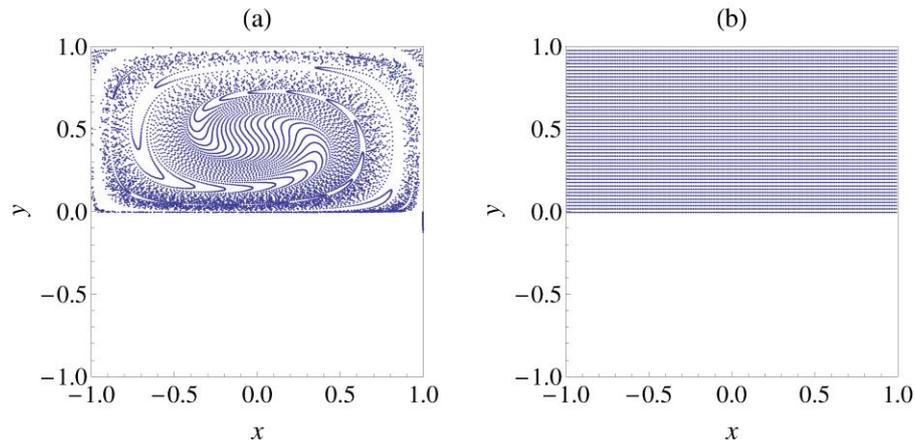

Fig 16 Location of the tracer particles after 5 unit cells for (a) circlet geometry (b) serpentine geometry. Initially all the particles are located in the upper half. The system parameters are: $\text{Re} = 15, \lambda = 1, k = 0$.

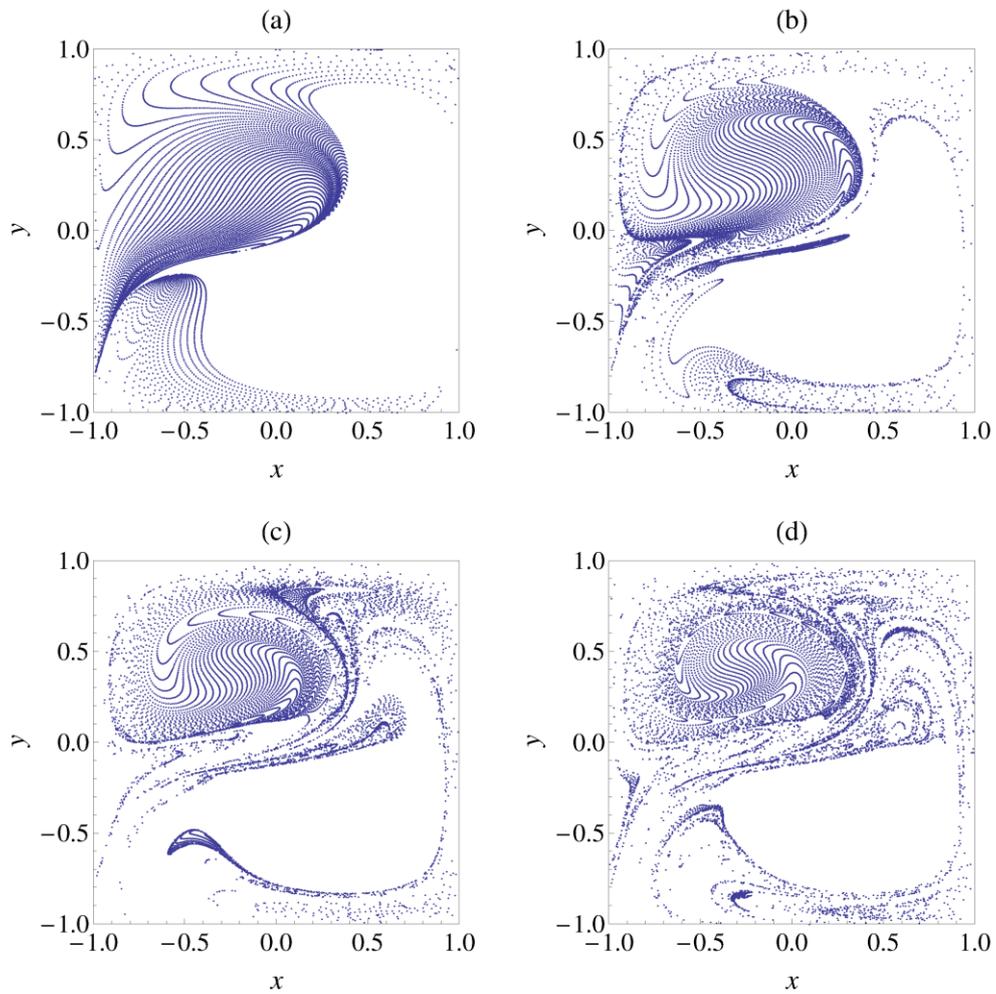

Fig 17 Location of the tracer particles after (a) 1 (b) 3 (c) 6 (d) 10 unit cells for the circlet geometry. Initially all the particles are located in the upper half. The system parameters are: $\text{Re}=15, \lambda=1, k=0.2$

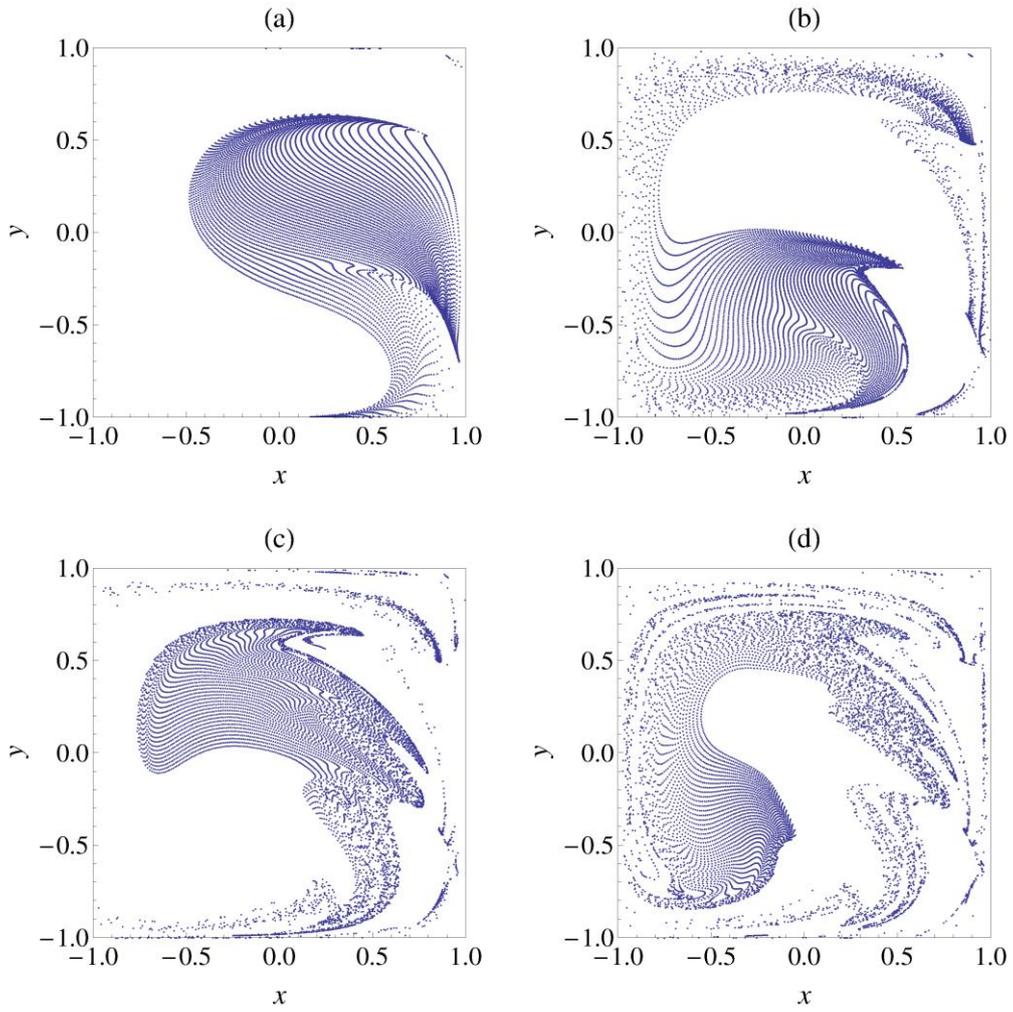

Fig 18 Location of the tracer particles after (a) 1 (b) 3 (c) 6 (d) 10 unit cells for the serpentine geometry. Initially all the particles are located in the upper half. The system parameters are: $\mathrm{Re}=15, \lambda=1, k=0.2$

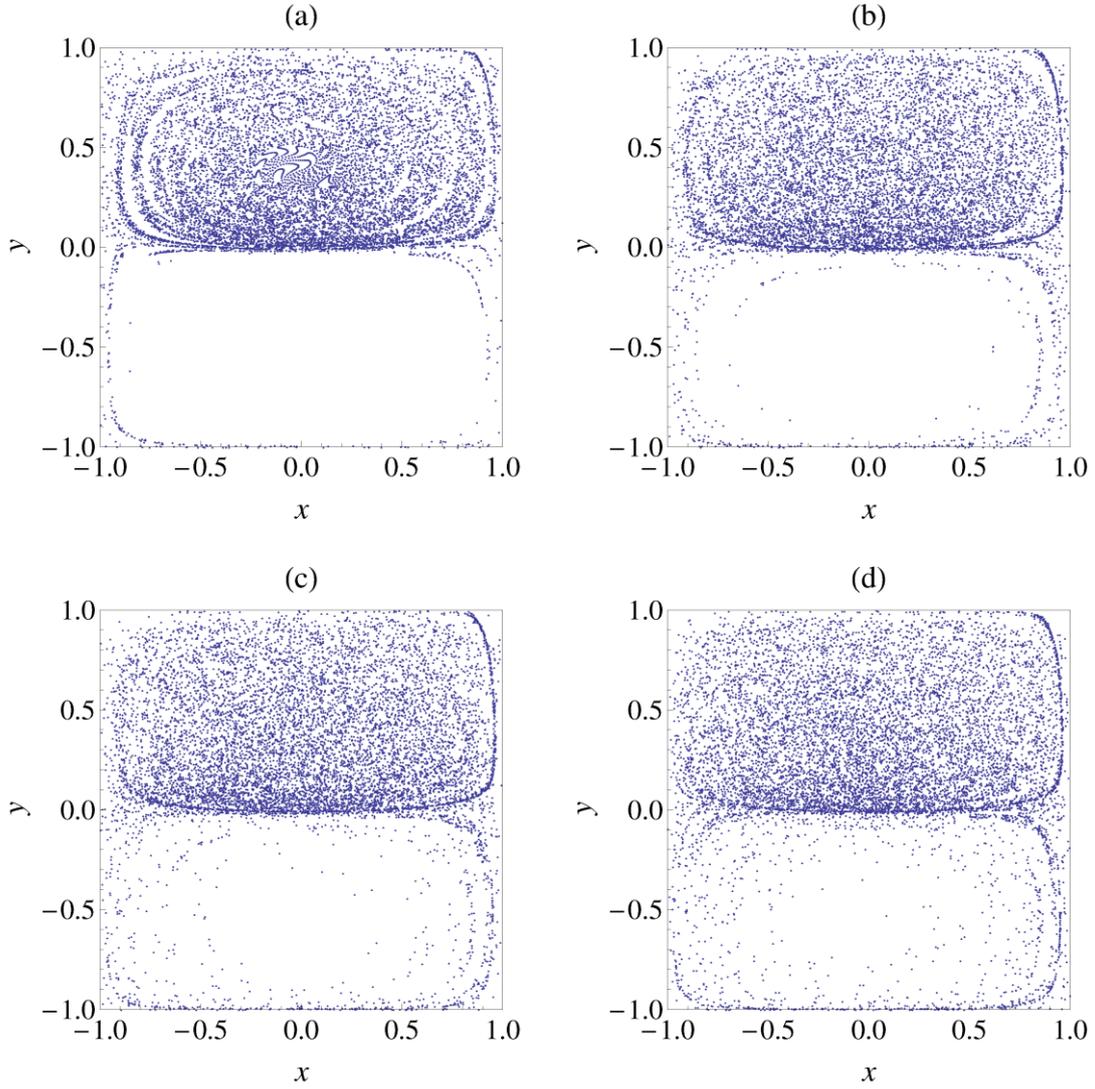

Fig 19 Location of the tracer particles after (a) 1 (b) 3 (c) 6 (d) 10 unit cells for the serpentine geometry. Initially all the particles are located in the upper half. The system parameters are: $\text{Re}=15, \lambda=3, k=0.2$

## VI. CONCLUDING REMARKS

In this work, we have presented and analyzed a novel micromixer design strategy for curved channels using slip surfaces. Our design exploited the 'movement of the separatrix' mechanism to cause chaotic mixing in curved channels. Until now, there was no discernible physical mechanism that could induce asymmetry about the separatrix in Dean flow. We demonstrated that it is easily obtained by making the axial velocity profile asymmetric. This work paves the way for designing a new of class of mixers in curved channels based on this idea. We also carried out a sweep of the parameter space, examining the effect of slip length and aspect ratio on chaotic mixing. Increased mixing in separatrix flows for wider channels has not been reported previously in the literature, despite the fact that separatrix flows have been studied for several different geometries.

Further work can focus on a study of the properties of the underlying symplectic maps, using dynamical systems theory. Since, chaotic mixing is essentially a kinematic phenomenon, explanations relying on just the dynamical features of the flow field (i.e. the physical forces) are not sufficient to completely understand the behavior of the system. Thus, examining the response to perturbations of the underlying maps studied in this paper can shed more insight on the variation of mixing efficiency for different parameters as well as provide a way for controlling the chaos. Studying the response to perturbations of the underlying maps (Linked Twist Maps with a separatrix) with respect to the aspect ratio in rectangular domains is a particularly interesting line of enquiry. Most practical applications impose constraints on the possible geometries which may be used. Using the understanding developed in this paper as a guide, the parameter space may thus be identified which would lead to most efficient mixing subject to the relevant constraints. Further work can then focus on relaxing the assumptions on the flow field imposed in the present model and study the mixing in more realistic flow fields.

While the results obtained in this study are independent of the mechanism of the slip, the optimization of the mixer with different slip mechanisms may yield slightly different results. For instance, it has been observed that wider channels have a greater effective slip length when superhydrophobic surfaces are used[26]. This compliments our results on wider channels as increasing slip in wider channels will lead to even better mixing than predicted here. We further note that geometry modifications such as herringbones, used in previous passive chaotic mixers, increase the pressure drop required to sustain the flow rate. However, since slip leads to drag reduction, the pressure drop will actually be less for the proposed slip-based chaotic mixer. Further, investigations are needed to quantify the reduced drag for curved channels with slippery surfaces. This added advantage should facilitate integration of the design to lab on chip devices.

**APPENDIX: SOLUTION OF THE VELOCITY FIELD**

Flow field in a curved channel of circular cross section was first determined by Dean[6], while Cuming[7] extended the solution to a channel of rectangular cross section. To analyze the chaotic mixer, we derive the velocity field in a curved channel with slip boundary conditions at either the top or bottom walls. The solution procedure closely follows Garg et al[22] and is briefly outlined here. We solve for the velocity field in half cell 1 of the circlet geometry i.e. slip at the top wall and centrifugal force directed along the positive x-direction. The velocity field for half cell 2 i.e. with slip at the bottom wall, of the circlet geometry is obtained using the transformation $y \rightarrow -y$, while of the serpentine geometry using $y \rightarrow -y, x \rightarrow -x$.

The continuity and Navier-Stokes equations in the aforementioned curvilinear co-ordinate system (Fig 3) are:

$$\frac{\partial u}{\partial x} + \lambda \frac{\partial v}{\partial y} + \varepsilon \frac{u}{(1+x\varepsilon)} = 0 \tag{A1}$$

$$\text{Re}\left(u\frac{\partial u}{\partial x} + \lambda v\frac{\partial u}{\partial y} - \varepsilon \frac{w^2}{(1+x\varepsilon)}\right) = \frac{\partial P}{\partial x} + \lambda^2 \frac{\partial^2 u}{\partial y^2} - \lambda \frac{\partial^2 v}{\partial x \partial y} \tag{A2}$$

$$\text{Re}\left(u\frac{\partial v}{\partial x} + \lambda v\frac{\partial v}{\partial y}\right) = -\lambda \frac{\partial P}{\partial y} + \frac{\partial^2 v}{\partial x^2} - \lambda \frac{\partial^2 u}{\partial y \partial x} + \frac{\varepsilon}{(1+x\varepsilon)}\left(\frac{\partial v}{\partial x} - \lambda \frac{\partial u}{\partial y}\right) \tag{A3}$$

$$\text{Re}\left(u\frac{\partial w}{\partial x} + \lambda v\frac{\partial w}{\partial y} + \varepsilon \frac{wu}{(1+x\varepsilon)}\right) = \frac{1}{(1+x\varepsilon)} + \frac{\partial^2 w}{\partial x^2} + \lambda^2 \frac{\partial^2 w}{\partial y^2}$$
$$+ \frac{\varepsilon}{(1+x\varepsilon)}\frac{\partial w}{\partial x} - \varepsilon^2 \frac{w}{(1+x\varepsilon)^2} \tag{A4}$$

These equations are subject to no-slip boundary conditions at the walls, excepting the top wall for half cell 1, where the Navier slip condition is used:

| | |
|---|---|
| $\{u,v,w\} = 0$ | at $y = -1$ for $-1 \leq x \leq 1$ |
| $\{u,v,w\} = 0$ | at $x = \pm 1$ for $-1 \leq y \leq 1$ |
| $v = 0$ | at $y = 1$ for $-1 \leq x \leq 1$ |
| $\{u,w\} = k(\hat{n}.\bar{\bar{\sigma}} - (\hat{n}.\hat{n}.\bar{\bar{\sigma}})\hat{n})$ | at $y = +1$ for $-1 \leq x \leq 1$ |
| $\{u,v,w\} = 0$ | at $x = \pm 1$ for $-1 \leq y \leq 1$ |

To obtain an approximate asymptotic solution for the limit of small curvature ratio and Reynolds number, we use the domain perturbation method[42,43]. The unknown variables (velocity and pressure) are expanded in an asymptotic series in the curvature ratio of the channel $(\varepsilon)$.

$$u(x,y;\varepsilon) = u_0(x,y) + \varepsilon u_1(x,y) + \varepsilon^2 u_2(x,y) + O(\varepsilon^3) \tag{A5}$$

$$v(x,y;\varepsilon) = v_0(x,y) + \varepsilon v_1(x,y) + \varepsilon^2 v_2(x,y) + O(\varepsilon^3) \tag{A6}$$

$$w(x,y;\varepsilon) = w_0(x,y) + \varepsilon w_1(x,y) + \varepsilon^2 w_2(x,y) + O(\varepsilon^3) \tag{A7}$$

Substituting (A5)-(A7) in (A1)-(A4) and equating terms of equal orders of $\varepsilon$, we obtain a system of equations which may be solved at each order one after the other. Using continuity, we define a stream function, $\psi$,:

$$u_1 = \lambda \frac{\partial \psi}{\partial y} \qquad v_1 = -\frac{\partial \psi}{\partial x}$$

Hence, the final equations that are to be solved are:

$$\frac{\partial^2 w_0}{\partial x^2} + \lambda^2 \frac{\partial^2 w_0}{\partial y^2} = -1 \tag{A7}$$

$$\frac{\partial^4 \psi}{\partial x^4} + 2\lambda^2 \frac{\partial^4 \psi}{\partial x^2 \partial y^2} + \lambda^4 \frac{\partial^4 \psi}{\partial y^4} = -\lambda \operatorname{Re} \frac{dw_o^2}{dy} \tag{A8}$$

The RHS of (A8) represents the centrifugal force. From this, we can see that the location of the maxima of the centrifugal force depends on the location of the maxima of the axial velocity. The above equations are solved using concepts from Linear Operator Theory[44,45]. The details of the method used can be found in Garg et al[22].

The final expression for the velocity field obtained is as follows:

$$w_0 = \sum_{r=0}^{\infty} \frac{16(-1)^r}{\pi^3 (2r+1)^3} \left( A_r \cosh\left(\frac{(2r+1)\pi y}{2\lambda}\right) + B_r \sinh\left(\frac{(2r+1)\pi y}{2\lambda}\right) + 1 \right) \cos\left(\frac{(2r+1)\pi x}{2}\right) \tag{A9}$$

$$\psi = \sum_{m=1}^{\infty} (C_1 \cosh(\omega_m x) + C_2 x \sinh(\omega_m x)) \phi_y$$

$$+ \sum_{n=1}^{\infty} \left( C_3 \cosh\left(\frac{\omega_n y}{\lambda}\right) + C_4 \sinh\left(\frac{\omega_n y}{\lambda}\right) + C_5 y \cosh\left(\frac{\omega_n y}{\lambda}\right) + C_6 y \sinh\left(\frac{\omega_n y}{\lambda}\right) + P(y) \right) \phi_x \tag{A10}$$

$$\phi_x = \cos\left(\frac{(2n-1)\pi x}{2}\right) \qquad \phi_y = \sin(m\pi y)$$

$$\omega_n = \frac{(2n-1)\pi}{2} \qquad \omega_m = m\pi\lambda$$

for $n = 1,2,3...$ \qquad for $m = 1,2,3...$

$P(y)$ is the particular solution of the fourth order ODE in $y$, which is obtained after projecting Eqn. A8 along $x$ [22]. The constants of integration are obtained using the boundary conditions.

Note that only the first term $(r = 0)$ in the expansion of $w_0$ (A9) was used while solving Eqn. A8.